# A Mini Review on Tumor Organoid-on-a-Chip Technologies in Personalized Oncology

Keyvan Alavi [1*]

[1*] Department of Biomedical Engineering, Science and Research Branch, Islamic Azad University, Tehran, Iran

email: Keivan.alavi@gmail.com - k.alavi@iau.ir

## Abstract

Tumor organoid-on-a-chip platforms represent a cutting-edge fusion of patient-derived organoids with microfluidic technologies, offering unprecedented capabilities for personalized cancer research. These systems overcome limitations of conventional models by enabling precise control over the tumor microenvironment, including nutrient gradients, fluid flow, and immune interactions. Tumor organoids recapitulate patient-specific tumor heterogeneity and genetic landscapes, while microfluidic chips provide dynamic perfusion and mechanical stimuli, enhancing physiological relevance. Together, they facilitate advanced applications such as high-throughput drug screening, immunotherapy testing, and metastasis modeling, showing superior predictive power for clinical outcomes. Despite challenges in standardization, scalability, and integration of complex tumor components, ongoing advances in hydrogel engineering, automation, and artificial intelligence are poised to accelerate their clinical translation. This review highlights current technologies, applications, and future directions of tumor organoid-on-a-chip systems, emphasizing their transformative potential in precision oncology.

**Keywords: Tumor Organoids, Microfluidics, Personalized Cancer Therapy, Tumor Microenvironment**

## 1. Introduction

One of the most significant challenges in cancer therapy is addressing tumor heterogeneity—the presence of genetically and phenotypically diverse subpopulations of cancer cells within a single tumor and across different tumor sites. This heterogeneity is a major driver of drug resistance, disease progression, and treatment failure, as distinct tumor cell populations may respond differently to therapeutic agents[4, 5]. Despite extensive research, traditional cancer models, such as two-dimensional (2D) cell cultures and animal models, offer limited capacity to capture this complexity. 2D cell cultures lack the three-dimensional (3D) architecture, cellular diversity, and dynamic microenvironment of tumors in the human body. Consequently, they fail to reproduce key features such as oxygen gradients, nutrient diffusion, and mechanical forces, leading to poor predictive power in drug screening[6]. Similarly, animal models, particularly murine xenografts, face inherent limitations due to species differences in immune responses, metabolism, and tumor biology[7]. Furthermore, these models typically rely on well-established cancer cell lines rather than patient-specific tumors, limiting their relevance for personalized therapy. Together, these limitations underscore the urgent need for advanced models that can more accurately replicate human tumor complexity and improve the predictability of therapeutic outcomes. In response to these limitations, 3D tumor organoids have emerged as powerful in vitro models that more closely resemble the structural and functional characteristics of human tumors. Derived from patient tumor tissues or stem cells, these self-organized mini-tumors retain key features such as tumor heterogeneity, genetic mutations, and cell–cell and cell–matrix interactions[8, 9]. Importantly, tumor organoids offer a patient-specific platform for studying cancer biology and testing therapeutic responses, making them attractive tools for personalized medicine[10]. In parallel, organ-on-a-chip technologies have gained increasing attention for their ability to recapitulate physiological microenvironments using microfluidic systems. These platforms allow precise control over fluid flow, mechanical cues, and chemical gradients, mimicking dynamic in vivo conditions such as nutrient perfusion, shear stress, and immune interactions[11]. By merging tumor organoids with organ-on-a-chip systems, researchers are now developing next-generation platforms that combine the biological complexity of patient-derived organoids with the environmental control of microfluidics. This hybrid approach offers new opportunities to advance personalized cancer treatment by enabling more accurate drug screening, mechanistic studies, and predictive modeling of therapeutic responses[8, 12]. This review focuses on recent developments in tumor organoid-on-a-chip platforms, highlighting their design, applications, and potential to transform cancer research and precision oncology. A comparison of traditional tumor models, organoids, and organoid-on-a-chip platforms is summarized in Table 1, based on previous studies[13-16].

Table 1 Comparison of conventional and emerging tumor models, summarizing their key features, advantages, and limitations in cancer research

| Feature | 2D Cell Cultures | Animal Models | Tumor Organoids | Tumor Organoid-on-a-Chip |
|---|---|---|---|---|
| Architecture | Flat monolayer | 3D (in vivo) | 3D self-organized | 3D with dynamic microenvironment |
| Tumor Microenvironment | Absent | Partial (species-specific) | Limited (no vasculature/immune cells) | Highly controllable and customizable |
| Patient Specificity | Low (immortalized lines) | Low to moderate (standardized xenografts) | High (patient-derived) | High (patient-derived + customizable TME) |
| Heterogeneity Representation | Poor | Moderate | Good | Excellent |
| Drug Screening Utility | Low predictive power | Moderate predictive power | Good predictive power | High predictive power |
| Immune Component | Absent | Animal immune system (non-human) | Typically absent | Co-culture with human immune cells possible |
| Perfusion & Flow Simulation | Absent | Present (natural) | Limited (diffusion only) | Present (controlled via microfluidics) |
| Reproducibility | High | Moderate | Variable | Variable (improving with standardization efforts) |
| Ethical Concerns | Minimal | High (animal | Low | Low |
|  |  | welfare issues) |  |  |
| Scalability & Throughput | High | Low | Moderate | Moderate (improving with automation) |

Beyond comparing features of existing tumor models (Table 1), it is also essential to visualize the historical progress of tumor organoid technologies, their integration with other experimental models, and emerging applications. Figure 1 provides a timeline of tumor organoid development, summarizes the available cancer modeling platforms, and highlights the integration of organoid-on-a-chip technologies for advanced immunotherapy testing.

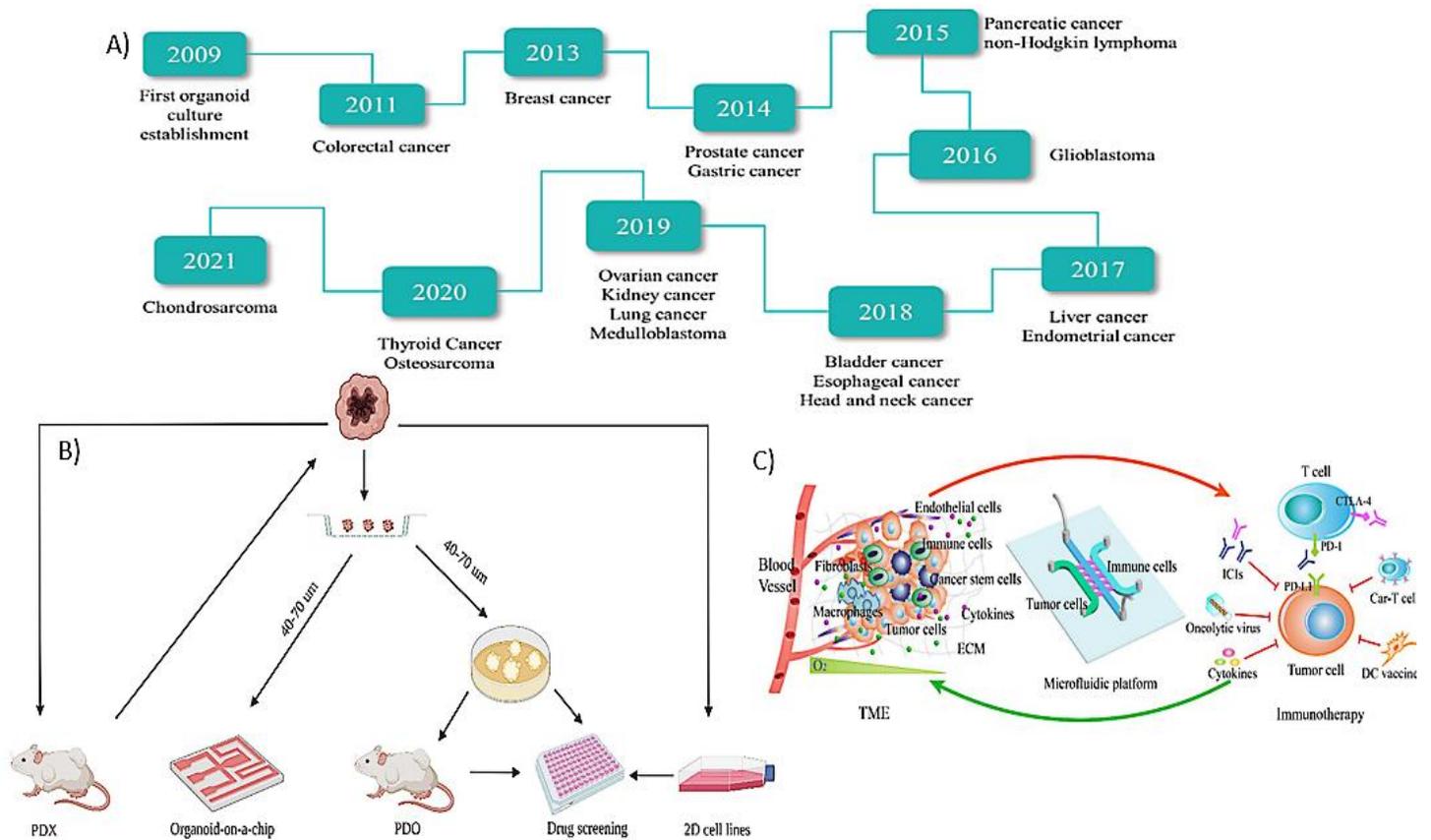

**Figure 1(A)** Timeline of key milestones in the development of tumor organoid technologies, highlighting the first establishment of organoid cultures and subsequent advancements across various cancer types[1]. **(B)** Schematic overview of in vitro cancer modeling platforms for drug screening. Patient-derived tumor samples can be processed into patient-derived organoids (PDOs), organoid-on-a-chip systems, patient-derived xenografts (PDX), or 2D cell lines for preclinical testing[2]. **(C)** Engineered tumor organoid-on-a-chip platforms enable precise modeling of the tumor microenvironment (TME), integrating tumor cells, stromal cells, immune components, and microfluidic perfusion. These systems are used for investigating cancer immunotherapies, including immune checkpoint inhibitors (ICIs), CAR-T cells, oncolytic viruses, and dendritic cell (DC) vaccines[3].

## 2. Tumor Organoids: Current State & Potential

Tumor organoids are 3D in vitro culture systems derived directly from patient tumor tissues[17]. These self-organizing structures closely mimic the genetic[13], histological, and functional characteristics of the original tumors, including key features such as tumor heterogeneity, mutational landscapes, and cellular architecture. Unlike traditional 2D cell lines, tumor organoids maintain patient-specific traits, making them valuable models for studying tumor biology and therapeutic responses on an individualized basis. One of the primary advantages of tumor organoids is their ability to support long-term culture while preserving the genetic stability and phenotypic diversity of the original tumor. This allows researchers to expand tumor organoids over multiple passages for experimental studies without losing critical tumor-specific attributes. Furthermore, their patient-specific nature enables direct evaluation of drug sensitivity and resistance, paving the way for personalized cancer treatment strategies. The generation of tumor organoids typically involves isolating cancer cells from patient biopsy or surgical specimens, followed by their encapsulation within hydrogel-based matrices that mimic the extracellular matrix (ECM). Common hydrogels used for this purpose include Matrigel (a basement membrane extract), collagen, and synthetic alternatives such as polyethylene glycol (PEG)-based hydrogels. These matrices provide essential structural support and biochemical cues that promote cell proliferation, differentiation, and self-organization into 3D organoids. In addition to the physical scaffold, the culture medium is supplemented with specific growth factors and signaling molecules that recreate the tumor's native niche and promote organoid formation. These factors vary depending on tumor type but generally target pathways such as Wnt, EGF, and TGF-β. Tumor organoid models have been successfully developed for a wide range of cancers[9], including colorectal, breast, pancreatic, and lung cancers. These models are now widely used in basic cancer research, drug screening, and the development of personalized therapeutic regimens. Despite their significant advantages, tumor organoids are not without limitations. One major drawback is their lack of vasculature, immune components, and mechanical stimuli, all of which play critical roles in tumor progression and therapy responses in vivo. Furthermore, as organoids grow larger, they may suffer from poor nutrient and oxygen diffusion, leading to central necrosis and restricted long-term viability. These limitations motivate the integration of tumor organoids with organ-on-a-chip platforms, where microfluidic systems can provide controlled fluid flow, nutrient perfusion, and mechanical stimulation to better replicate the in vivo tumor microenvironment. Recent studies have employed tumor organoid models for personalized cancer research and therapy development. Table 2 highlights organoid-based applications across various cancer types, including drug discovery, chemotherapy response prediction, and translational research.

*Table 2* Recent representative studies utilizing patient-derived organoid models for personalized cancer therapy. The table summarizes key applications across different cancer types, including drug screening, chemotherapy response prediction, and precision oncology research.

| No. | Title | Authors | Year | Focus/Key Insights | Reference |
|---|---|---|---|---|---|
| 1 | Organoid Technology as a Next Generation Platform for Evaluating Drug Discovery System | Mahale, B.M., et al. | 2025 | Use of tumor organoids for drug discovery and personalized treatment testing | 18 |
| 2 | Current Landscape of Preclinical Models for Pediatric Gliomas | Yadav, M., et al. | 2025 | Pediatric glioma organoid models for preclinical research and personalized therapies | 19 |
| 3 | Recent Advances in Colorectal Cancer: From Molecular Research to Treatment | Qi, G.X., et al. | 2025 | Tumor organoids for colorectal cancer drug testing and individualized treatments | 20 |
| 4 | Patient-derived organoids: Advancing research on bioactive natural compounds in lung cancer | Chen, X., Lin, X. | 2025 | Lung cancer organoids for drug screening and personalized therapy | 21 |
| 5 | Pancreatic Organoids as Cancer Avatars for Personalized Medicine | Gout, J., et al. | 2025 | Pancreatic cancer organoids for precision oncology | 22 |
| 6 | Exploring Personalized Prediction of Clinical Chemotherapy Efficacy and | Du, L., et al. | 2025 | 3D bioprinted gastric cancer organoids for chemotherapy prediction | 23 |

| | Revealing Tumor Heterogeneity Using Patient-Derived 3D Bioprinting Gastric Cancer Models | | | | |
|---|---|---|---|---|---|
| 7 | An Organoid - Guided Platform for Ovarian Cancer: Enabling Prediction of Patients' Chemotherapy Response | Wang, L., et al. | 2025 | Ovarian cancer organoids for chemotherapy response prediction | 24 |
| 8 | Anti-cancer drug sensitivity testing and preclinical evaluation of the anti-cancer potential of WEE1 inhibitor in triple-negative breast cancer patient-derived organoids and xenograft models | Ryu, S., et al. | 2025 | Preclinical drug testing using breast cancer organoids | 25 |
| 9 | Endocrine Cancer Organoids in Translational Research | Chen, D., et al. | 2025 | Endocrine cancer organoids for drug testing and translational research | 26 |
| 10 | Next-Gen Models of Gallbladder Carcinoma: Towards Precision Medicine | Sahoo, O.S., et al. | 2025 | Gallbladder organoid systems advancing precision oncology | 27 |

## 3. Microfluidic Chips: Design & Functionality

Microfluidic chip technology involves the manipulation of fluids at the microscale through precisely engineered networks of microchannels[28]. These chips are typically fabricated from biocompatible materials such as polydimethylsiloxane (PDMS), glass, or thermoplastics, and are designed to enable the controlled flow of fluids within channels that can range from tens to hundreds of micrometers in diameter[29]. By regulating fluid flow, microfluidic systems can create well-defined chemical gradients, which are critical for studying how cells respond to varying concentrations of drugs, nutrients, or signaling molecules. Importantly, microfluidic chips allow the simulation of key physiological processes within the tumor microenvironment. Through the design of interconnected channels and chambers, these devices can mimic vascular perfusion, allowing nutrient and oxygen transport similar to that in blood vessels. Additionally, microfluidic chips can generate shear stress and interstitial flow, which are known to influence tumor progression, invasion, and drug transport. These features enable researchers to study cancer cells under dynamic, flow-based conditions that more closely resemble the in vivo environment. One of the most powerful features of microfluidic chip platforms is their ability to precisely regulate the tumor microenvironment. By controlling oxygen levels, these systems can recreate hypoxia, a hallmark of solid tumors that profoundly affects cancer metabolism, drug resistance, and aggressiveness. Similarly, nutrient gradients can be established within the chip to mimic the metabolic heterogeneity observed in tumors, enabling the study of cancer cell adaptation to nutrient-deprived or nutrient-rich zones. Furthermore, microfluidic chips facilitate dynamic drug perfusion studies, allowing researchers to investigate drug delivery and efficacy under flow conditions that simulate those in blood vessels and tissues[30]. Unlike static culture systems, these chips can model drug transport kinetics, penetration depths, and accumulation within tumors in real time. This capability is particularly relevant for evaluating the pharmacokinetics of chemotherapeutic agents and testing new drug formulations. By replicating these critical aspects of the tumor microenvironment, microfluidic chips provide a versatile platform for advanced cancer research and therapeutic screening[31].

## 4. Tumor Organoid-on-a-Chip Platforms: Current Technologies

The integration of tumor organoids with microfluidic chips has led to the development of tumor organoid-on-a-chip platforms, which offer substantial advantages over conventional organoid culture methods. By embedding organoids within the microfluidic environment, researchers can achieve enhanced nutrient and waste exchange through controlled fluid flow, overcoming diffusion limitations commonly observed in static 3D cultures[3]. This improved perfusion not only supports sustained organoid viability and growth but also enables better replication of in vivo-like nutrient and oxygen gradients[32]. Additionally, the microfluidic format allows for precise environmental control and real-time monitoring of organoid behavior under various experimental conditions. Researchers can introduce drugs, immune cells, or mechanical stimuli in a dynamic and reproducible manner while observing morphological and functional changes via integrated imaging systems[33, 34]. These features make organoid-on-a-chip systems highly versatile platforms for studying cancer progression, drug responses, and mechanisms of resistance at single-organoid or high-throughput scales[3]. The successful fabrication of tumor organoid-on-a-chip devices relies on the careful selection of materials that support both chip function and biological compatibility. Common materials include PDMS, known for its gas permeability and ease of fabrication; glass, valued for optical transparency and chemical stability; and various thermoplastics that offer improved scalability for mass production. While PDMS remains popular in academic research due to its flexibility and low cost, its tendency to absorb hydrophobic molecules raises concerns for drug screening applications. Therefore, material selection must balance biocompatibility, optical properties, and scalability based on the intended application. Within these chips, tumor organoids are typically embedded in hydrogel matrices that mimic the ECM. Common hydrogels include Matrigel, collagen, and synthetic polymers such as polyethylene glycol (PEG)-based hydrogels. These materials provide essential biochemical and mechanical cues that guide organoid development and function. Importantly, matrix stiffness, degradation rates, and bioactivity must be carefully tuned to match the mechanical properties of native tissues and support long-term viability. Advances in hydrogel engineering now allow for precise control of these parameters, enabling the creation of more physiologically relevant tumor models within organoid-on-a-chip systems[34, 35]. To further enhance physiological relevance, many tumor organoid-on-a-chip platforms incorporate co-culture systems that include additional cell types present in the tumor microenvironment (TME). These may involve immune cells, such as T cells or macrophages, to study immune-tumor interactions; fibroblasts to model stromal components; or endothelial cells to simulate vascularization and angiogenesis[32, 34, 35]. The ability to control the spatial arrangement and interaction between these cell types within microfluidic chips allows for detailed investigation of complex cellular crosstalk in cancer progression and therapy resistance. Several studies have demonstrated the utility of such co-culture systems. For example, immune-enhanced organoid chips have been used to evaluate immunotherapy responses by tracking cytotoxic T-cell activity against tumor organoids under flow conditions. Similarly, models incorporating endothelial cells have been employed to study tumor angiogenesis and drug transport across vascular barriers[33, 36]. These advanced platforms represent a significant step forward in replicating the complexity of in vivo tumors within an accessible, controllable in vitro setting.

## 5. Applications in Personalized Cancer Therapy

Tumor organoid-on-a-chip platforms offer a powerful approach for high-throughput drug screening and [37], addressing the urgent need for more predictive preclinical cancer models. By combining patient-derived organoids with microfluidic chips, researchers can test the efficacy of various anti-cancer drugs under physiologically relevant conditions, including dynamic fluid flow and 3D architecture[38]. These systems enable precise control over drug concentration, exposure duration, and perfusion patterns, which are critical for accurately assessing drug responses. Notably, several studies have demonstrated the utility of these platforms in specific cancer types[7, 11]. In colorectal cancer, organoid-on-a-chip systems have been employed to evaluate chemotherapy regimens such as 5-fluorouracil and oxaliplatin, showing improved correlation with clinical outcomes compared to traditional models[39]. Similarly, pancreatic cancer models have been developed to assess responses to gemcitabine and other chemotherapeutics, revealing patient-specific drug sensitivities that could guide personalized treatment plans. These applications highlight the potential of organoid-on-a-chip technologies to revolutionize drug development and improve treatment selection for cancer patients. In addition to chemotherapy screening, tumor organoid-on-a-chip platforms have shown significant promise in the realm of immunotherapy testing. By incorporating immune components such as T cells and natural killer (NK) cells into the chip environment, researchers can model immune-tumor interactions with greater precision than ever before. This enables the study of key processes such as immune cell infiltration, cytotoxicity, and tumor immune evasion within a patient-specific context. One of the most exciting applications of these platforms lies in checkpoint blockade therapy, a rapidly growing class of cancer treatments targeting immune checkpoint pathways such as PD-1/PD-L1 and CTLA-4. Tumor organoid-on-a-chip systems can be engineered to co-culture tumor organoids with autologous or engineered immune cells under flow conditions, allowing for functional testing of [3]. Such platforms have been utilized to predict individual patient responses to immunotherapies, potentially guiding clinical decisions and improving treatment outcomes. Tumor organoid-on-a-chip systems also provide unique opportunities for studying cancer metastasis and invasion, two of the most lethal aspects of cancer progression. By leveraging the ability of microfluidic chips to generate chemical gradients and flow patterns, researchers can model how cancer cells migrate and invade surrounding tissues under conditions that mimic the tumor microenvironment[37]. Recent studies have used these platforms to track cancer cell migration and epithelial-to-mesenchymal transition (EMT), processes that are central to metastatic dissemination. For example, breast cancer organoid-on-a-chip models have been used to examine how mechanical cues and matrix stiffness influence invasive behavior[40]. Additionally, platforms that simulate lymphatic or vascular channels enable the study of intravasation and extravasation, key steps in the metastatic cascade. These tools are invaluable for uncovering the mechanisms underlying metastasis and for testing anti-metastatic therapies in a controlled, reproducible setting. Beyond preclinical research, tumor organoid-on-a-chip platforms hold strong potential as predictive tools for clinical decision-making[41]. Several proof-of-concept studies have shown that patient-derived organoid chips can predict therapeutic responses that closely match actual clinical outcomes. These findings suggest that such platforms could be used to personalize treatment plans by identifying the most effective therapies for individual patients before initiating treatment. However, despite their promise, several challenges remain before these technologies can be fully implemented in clinical settings. Issues such as scalability, cost, and the time required to generate patient-specific organoid chips may limit widespread adoption in routine oncology practice. Additionally, regulatory frameworks for approving personalized organoid-based testing are still under development. Nonetheless, continued advances in automation, microfabrication, and biomanufacturing could help address these barriers and bring tumor organoid-on-a-chip systems closer to clinical integration.

## 6. challenges & Future Directions

Despite their rapid development and significant potential, tumor organoid-on-a-chip platforms face several critical challenges that must be addressed to enable broader adoption in research and clinical settings. One of the foremost issues is the reproducibility and standardization of both organoid generation and chip fabrication[42, 43]. Variability in patient-derived samples, hydrogel properties, and microfluidic chip designs can lead to inconsistent results, limiting cross-study comparisons and slowing progress toward clinical applications. Another major challenge lies in engineering functional vasculature and immune components within these systems. While some progress has been made in incorporating endothelial cells and immune cells into chips, fully replicating the dynamic, multicellular interactions of the tumor microenvironment remain technically difficult. Additionally, the cost and complexity of fabrication present practical barriers, particularly for high-throughput drug screening and personalized medicine. Advanced microfabrication techniques and materials, though increasingly sophisticated, often require specialized equipment and expertise, which can limit widespread accessibility. Looking forward, several promising strategies are emerging to overcome these barriers and further advance tumor organoid-on-a-chip technologies. One key direction is the integration of artificial intelligence (AI) and machine learning with these platforms to enable automated image analysis, pattern recognition, and predictive modeling of drug responses. Such tools could greatly enhance the ability to extract meaningful data from complex organoid-chip systems and accelerate personalized therapy development. Another exciting avenue involves incorporating these platforms into personalized treatment pipelines, where patient-derived organoid chips are used for rapid drug screening to inform clinical decision-making. Advances in biomanufacturing and automation may help reduce production time and costs, making these platforms more feasible for clinical use. However, challenges related to regulatory approval, scalability, and quality control must be carefully addressed to enable clinical translation. As the field continues to evolve, interdisciplinary collaboration between bioengineers, clinicians, regulatory bodies, and data scientists will be essential to fully realize the transformative potential of tumor organoid-on-a-chip systems in personalized cancer care.

## 7. Conclusion

Tumor organoid-on-a-chip systems represent a significant advancement in the field of personalized oncology, offering a unique combination of biological complexity and environmental control that surpasses traditional cancer models. By integrating patient-derived tumor organoids with microfluidic technologies, these platforms provide a powerful tool for recapitulating the tumor microenvironment, enabling more accurate drug screening, immunotherapy testing, and mechanistic studies. Importantly, they hold great promise for bridging the gap between laboratory research and clinical application, potentially guiding individualized treatment decisions and accelerating the development of precision therapies. To fully realize their potential, however, continued interdisciplinary collaboration will be essential, bringing together expertise from materials science, microfluidics, cancer biology, bioengineering, and computational modeling to address remaining challenges and drive clinical translation of these innovative technologies.

## Ethical Compliance Statement

I hereby confirm full compliance with established publication ethics and best practices. I affirm that this manuscript is solely my original work, has not been published elsewhere in any language, and is not under consideration by any other journal. I have avoided any form of figure manipulation, ensured the accuracy of data presented, and disclose no conflicts of interest. Furthermore, I confirm that all aspects of research ethics and integrity have been strictly followed.

## Conflicts of Interest

The author declares no conflicts of interest.

## Data availability

No primary research data, software, or code were generated or analyzed in the preparation of this review article. As this is a review based on previously published literature, no new data were created, and therefore, no datasets are associated with this work.

## Acknowledgments

I would like to express my heartfelt gratitude to my family for their unwavering support, encouragement, and understanding throughout the preparation of this work.